\documentclass[aps,physrev,reprint,superscriptaddress,longbibliography,nofootinbib,floatfix]{revtex4-2}
\usepackage{amsmath,amssymb,bm,graphicx}
\usepackage[colorlinks=true,linkcolor=blue,citecolor=blue,urlcolor=blue]{hyperref}

\newcommand{\dd}{\mathrm d}
\newcommand{\ii}{\mathrm i}
\newcommand{\Tc}{T_{c}}
\newcommand{\DeltaZ}{\Delta_0}
\newcommand{\xiz}{\xi}
\newcommand{\Js}{J_{s0}}

\newcommand{\G}{\mathcal G}
\newcommand{\Jcal}{\mathcal J}

\begin{document}

\title{Microscopic equivalence of the vortex-entry current and the depairing current in a superconducting thin-film strip}

\author{Takayuki Kubo}
\email{kubotaka@post.kek.jp}
\affiliation{High Energy Accelerator Research Organization (KEK), Tsukuba, Ibaraki 305-0801, Japan}
\affiliation{The Graduate University for Advanced Studies (Sokendai), Hayama, Kanagawa 240-0193, Japan}


\begin{abstract}
The vortex-entry current density \(J_{\rm v}\) of a superconducting strip is usually defined, within phenomenological Pearl--London theory, as the current density at which the edge barrier for vortex entry disappears. 
In that approach, \(J_{\rm v}\) depends on a short-distance core cutoff introduced by hand, and its temperature dependence cannot be determined within the same framework. 
To remove this cutoff ambiguity and determine the temperature dependence, one needs a microscopic calculation of the vortex-entry current. 
Nevertheless, such a microscopic calculation has never been carried out. 
Here, we formulate and solve this problem for an ideal homogeneous dirty-limit superconducting thin-film strip at zero applied field, with self-field effects neglected. 
Vortex entry is treated as the loss of local stability of the vortex-free current-carrying state. The calculation uses the fixed-current Gibbs functional of Usadel theory, which is valid over the full temperature range \(0<T<\Tc\), and examines both spatially uniform and nonuniform perturbations. 
The microscopic calculation shows that the condition for disappearance of the vortex-entry barrier is identical to the depairing condition. 
The central result is not merely that two current densities have the same value. The criterion for disappearance of the vortex-entry barrier and the depairing criterion are not independent conditions. Both identify the same loss of local stability of the vortex-free current-carrying state, namely, the same spinodal. Consequently, \(J_{\rm v}(T)=J_{\rm dp}(T)\) for all \(0<T<T_c\). This result determines the temperature dependence of \(J_{\rm v}\), removes the Pearl--London core-cutoff ambiguity, and establishes the microscopic equivalence of the vortex-entry and depairing current criteria.
\end{abstract}

\maketitle

\section{Introduction}

In phenomenological Pearl--London theory, vortex entry into a current-carrying superconducting thin-film strip is described in terms of an edge barrier. The vortex-entry current density \(J_{\rm v}\) is defined as the current density at which this barrier disappears. Consider a strip extending along the \(y\) direction and occupying \(-W/2\le x\le W/2\). The Pearl length is \(\Lambda=2\lambda^2/d\), where \(\lambda\) is the London penetration depth and \(d\) is the film thickness. In the narrow-strip limit \(\Lambda\gg W\), the self-field of the transport current can be neglected, and the bias sheet-current density is uniform.
For a point vortex at \({\bf r}_v=(X,0)\) and a given transport current \(I\), Pearl--London theory gives an energy profile \(G(X;I)\) as a function of the prescribed vortex position \(X\). 
The energy contains the London self-energy of the vortex and the work done by the transport current. 
For a vortex driven into the strip by the Lorentz force, the current term lowers the energy as the vortex moves inward from the edge~\cite{Likharev, Maksimova, Bartolf, Yamashita, Tafuri, Bulaevskii, Kubo_2023}. 
At low current, however, moving a vortex inward first increases \(G\), so the vortex-free state is protected by an entry barrier along this prescribed coordinate. 
The Pearl--London vortex-entry current is defined as the current at which this barrier disappears. 
This gives the well-known Pearl--London result for the vortex-entry current density~\cite{Tafuri,Bulaevskii,Kubo_2023}, \(J_{\rm v}^{\rm (L)}=I_{\rm v}^{\rm (L)}/Wd
=\phi_0/(\pi e\mu_0\Lambda\xi_{\rm cut}d)\), where \(e=2.718\ldots\) is Euler's number, not the elementary charge, and \(\xi_{\rm cut}\) is a short-distance cutoff of order the coherence length \(\xi\).
The Pearl--London result has two limitations. First, its value depends on the short-distance cutoff \(\xi_{\rm cut}\), which is introduced by hand. Second, the theory does not determine the temperature dependence of \(J_{\rm v}\) from its own equations.

Up to numerical factors, this result gives \(J_{\rm v}^{\rm (L)}\sim B_c/\mu_0\lambda\sim J_{\rm dp}\), since \(B_c\sim\phi_0/(\xi\lambda)\). Here, \(J_{\rm dp}\) is the depairing current density, above which the superflow can no longer sustain the superconducting condensate~\cite{Maki,Kupriyanov,Clem_Kogan,Kubo_2020, Kubo_erratum}. This observation raises a basic question: what is the relation between \(J_{\rm v}\) and \(J_{\rm dp}\)? Pearl--London theory, with its prescribed vortex coordinate and core cutoff, cannot answer this question unambiguously because the result depends on the short-distance cutoff $\xi_{\rm cut}$.

A cutoff-free treatment of vortex entry requires at least a Ginzburg--Landau (GL) description, although GL theory is valid only near \(T_c\).  
In this description, the barrier is not obtained by placing a London point vortex at a prescribed position.  Instead, one finds a saddle point of the current-biased GL functional. 
The barrier is the energy difference between this saddle point and the vortex-free state.
Such GL saddle points have been studied in mesoscopic disks and rings~\cite{Baelus}, at edges and surfaces~\cite{Vodolazov_2003}, and in current-biased two-dimensional films near the depairing
current~\cite{Vodolazov_2012}. 
These studies determine the saddle self-consistently from the current-biased GL functional, including both the order parameter and the current distribution.  
Near the depairing current, the saddle is not well described as a London point vortex.
Instead, it contains a localized region where the order parameter is suppressed. 
Thus a cutoff-free formulation naturally leads from a Pearl--London-like point-vortex calculation to a stability analysis of a current-biased functional.

The GL formulation removes the short-distance cutoff ambiguity, but it is valid only near \(\Tc\). It therefore determines the limiting behavior of \(J_{\rm v}(T)\) close to \(\Tc\), but not its temperature dependence over the full range \(0<T<\Tc\). A microscopic calculation is needed to determine \(J_{\rm v}(T)\) over the entire temperature range below \(\Tc\). Nevertheless, such a microscopic calculation has never been carried out.

This paper identifies the barrier-disappearance current within the dirty-limit Usadel theory, a microscopic theory of superconductivity that is valid at arbitrary temperatures in the range \(0<T<T_c\).
The logic is the same as in the GL formulation. 
One considers the fixed-current Gibbs functional \(\G\), written in terms of the gap \(\Delta\), the gauge-invariant superfluid momentum \(q\), and the quasiclassical Matsubara Green functions. 
At low current, the vortex-free current-carrying state is a local minimum of \(\G\). 
Vortex entry is blocked by a saddle point of the same functional, and the vortex-entry barrier is the energy difference between this saddle point and the vortex-free state. 
If the vortex-free state is stable against all allowed small perturbations, apart from the physically irrelevant zero mode corresponding to a spatially uniform phase shift, then \(\delta^2\G>0\) for every physically distinct nonzero perturbation.
The vortex-free state is then a strict local minimum of \(\G\). Any continuous vortex-entry path leaving this state must first pass through states of higher Gibbs energy, so the barrier cannot disappear while the vortex-free state remains locally stable.
As the current is increased, the barrier decreases. 
At the barrier-disappearance current, the saddle point merges with the vortex-free state. 
This is a saddle-node bifurcation: the local minimum and the saddle become the same stationary point, and the quadratic energy cost along one direction vanishes~\cite{Strogatz}. 
The current at which this happens is the cutoff-free microscopic counterpart of the Pearl--London vortex-entry current \(J_{\rm v}\). 
Therefore, it is not necessary to know the detailed shape or height of the finite barrier in order to determine \(J_{\rm v}\).
This is the same stability viewpoint used in GL~\cite{Transtrum, Gurevich_2023} and microscopic~\cite{Catelani, Kubo_2026} calculations of the superheating field.

The central result is conceptual rather than a numerical coincidence between two current scales. In the ideal strip, the disappearance of the vortex-entry barrier and depairing do not define two independent instabilities. They describe the same spinodal of the vortex-free current-carrying state.

\section{Microscopic theory of the vortex-entry current}

\subsection{Microscopic Gibbs functional} 

We consider the simplest microscopic reference problem: a homogeneous dirty superconducting thin-film strip at zero applied magnetic field, with current crowding, edge roughness, and magnetic self-field effects excluded. 
The strip is infinite in the longitudinal direction. In physical units, its width satisfies \(W\ll\Lambda\), where \(\Lambda\) is the Pearl length. 
Therefore the magnetic self-field of the transport current can be neglected.
In this ideal problem, the vortex-free state has a uniform gauge-invariant momentum. 
We follow the convention of Refs.~\cite{Gurevich_Kubo,Kubo_2020}. 
The gauge-invariant momentum is \({\bf q}=\nabla\chi+(2\pi/\phi_0){\bf A}\). 
With this sign convention, the supercurrent flows opposite to \({\bf q}\). 
All lengths and energies are measured in units of \(\xiz=\sqrt{\hbar D/(2\DeltaZ)}\) and the zero-temperature, zero-current gap \(\DeltaZ\), respectively. 
We define \({\bf Q}:=\xiz{\bf q}\), \(w:=W/\xiz\), \(\bar{\bf r}:={\bf r}/\xiz\), \(\bar\nabla:=\xiz\nabla\), \(\bar\Delta:=\Delta/\DeltaZ\), and the dimensionless Matsubara frequency \(\Omega_n:=\hbar\omega_n/\DeltaZ=2\pi\tau(n+1/2)\), where \(\tau:=k_BT/\DeltaZ\). 
The Matsubara Green functions are parametrized as \(g_n=\cos\theta_n\) and \(f_n=\sin\theta_n e^{\ii\chi}\).
Below, the bars are omitted, and \({\bf r}\), \(\nabla\), and \(\Delta\) denote the dimensionless coordinate, gradient, and gap, respectively.

The vortex-free state and its perturbations are described by the following self-consistent equations:
\begin{eqnarray}
&&
\nabla^2\theta_n
-|{\bf Q}|^2\sin\theta_n\cos\theta_n
-\Omega_n\sin\theta_n
+\Delta\cos\theta_n=0,
\label{eq:usadel}
\\
&&
\Delta\ln\frac{T}{\Tc}
+2\pi\tau\sum_{n=0}^{\infty}
\left(
\frac{\Delta}{\Omega_n}-\sin\theta_n
\right)=0,
\label{eq:gap}
\\
&&
\nabla\cdot\left[S({\bf r}){\bf Q}({\bf r})\right]=0,
\qquad
S:=2\pi \tau \sum_{n=0}^{\infty}\sin^2\theta_n .
\label{eq:phase}
\end{eqnarray}
They are the Usadel equation, the gap equation, and supercurrent conservation, respectively.  
The corresponding dimensionless current density is \({\bf J}/\Js=-(2/\sqrt{\pi})S{\bf Q}\), where
\(\Js:=B_{c0}/(\mu_0\lambda_0)\), \(B_{c0}\) is the zero-temperature thermodynamic critical field, and $\lambda_0$ is the zero-temperature, zero-current penetration depth.  
Equations~(\ref{eq:usadel})--(\ref{eq:phase}) follow from the stationarity conditions \(\delta F/\delta\theta_n=0\), \(\delta F/\delta\Delta=0\), and \(\delta F/\delta\chi=0\), respectively. 
The phase enters \(F\) only through the gauge-invariant momentum \({\bf Q}[\chi,{\bf A}]=\xi[\nabla\chi+(2\pi/\phi_0){\bf A}]\). 
Thus we write \(F[\{\theta_n\},\Delta,\chi]\), with \({\bf Q}\) understood as a functional of \(\chi\) and \({\bf A}\). 
The functional is
\begin{eqnarray}
&&
F[\{\theta_n\},\Delta,\chi]
 =
 \int \dd^2 r\,
 \biggl\{
 \Delta^2\ln\frac{T}{\Tc}
 +2\pi\tau\sum_{n=0}^{\infty}
 \Bigl[
 \frac{\Delta^2}{\Omega_n}
 \nonumber\\
&& +|\nabla\theta_n|^2
 -2\Delta\sin\theta_n
 +2\Omega_n(1-\cos\theta_n)
 +|{\bf Q}|^2\sin^2\theta_n
 \Bigr]
 \biggr\}.\nonumber \\
 \label{eq:free_energy}
\end{eqnarray}
At \(T=0\), the Matsubara sum is replaced by \(2\pi\tau\sum_{n\ge0}\to\int_0^\infty\dd\Omega\), after the standard regularization.

For an imposed positive dimensionless current-density magnitude
\(\Jcal_{\rm ext}\), the corresponding fixed-current Gibbs functional
for a strip segment of dimensionless area \(\mathcal A\) is
\begin{equation}
G[\{\theta_n\},\Delta,\chi]
=
F[\{\theta_n\},\Delta,\chi]
-\sqrt{\pi}\,\Jcal_{\rm ext}
\int_{\mathcal A}\dd^2 r\,Q_y .
\label{eq:gibbs_functional}
\end{equation}
Here, \(\Jcal_{\rm ext}\) is normalized by \(\Js\) and denotes the
positive magnitude of the imposed current density. The Legendre term
is conjugate to the spatially averaged longitudinal phase gradient.

For a spatially uniform vortex-free state, the gauge-invariant
momentum is uniform and points along the strip. Since the current
flows opposite to \({\bf Q}\), we denote by \(Q>0\) the momentum
magnitude and by
\(\Jcal(Q):=|{\bf J}/\Js|
=4\sqrt{\pi}\tau Q\sum_{n=0}^{\infty}\sin^2\theta_n(Q)\)
the positive dimensionless current density. For each \(Q\), the
spectral angles \(\theta_n(Q)\) and the gap \(\Delta(Q)\) are
determined self-consistently by the uniform Usadel equation
\(-Q^2\sin\theta_n\cos\theta_n-\Omega_n\sin\theta_n
+\Delta\cos\theta_n=0\), together with the gap
equation~\eqref{eq:gap}.
We define the free-energy and Gibbs-energy densities of this uniform
state by \({\mathcal F}(Q):=F/\mathcal A\) and
\({\mathcal G}(Q):=G/\mathcal A\), respectively.
Equation~\eqref{eq:gibbs_functional} then reduces to
\begin{equation}
{\mathcal G}(Q)
=
{\mathcal F}(Q)-\sqrt{\pi}\,\Jcal_{\rm ext}Q .
\label{eq:uniform_gibbs_density}
\end{equation}
With the present normalization,
\(d{\mathcal F}/dQ=\sqrt{\pi}\Jcal(Q)\). The implicit \(Q\)
dependence of \(\Delta\) and \(\theta_n\) gives no additional terms
because the Usadel and gap equations are satisfied. The stationarity
condition \(d{\mathcal G}/dQ=0\) therefore gives
\(\Jcal(Q)=\Jcal_{\rm ext}\).

\subsection{Uniform perturbations}

The microscopic stability problem is now formulated for the vortex-free current-carrying state. 
We first analyze spatially uniform perturbations and obtain a candidate stability limit. 
We then examine spatially nonuniform perturbations and show that, in the ideal zero-field strip with self-field effects neglected, they carry an additional positive stiffness and cannot become unstable before the uniform mode.

We first perturb the uniform state by a homogeneous superfluid-momentum
perturbation \(V\), a gap perturbation \(\eta\), and spectral-angle
responses \(\alpha_n\):
\begin{eqnarray}
&& Q\to Q+V, \label{uniform_V} \\
&& \Delta\to\Delta+\eta, \label{uniform_eta}\\
&& \theta_n\to\theta_n+\alpha_n .\label{uniform_alpha}
\end{eqnarray}
The fixed-current Gibbs energy \(\mathcal G\) differs from the
fixed-\(Q\) free-energy by a term linear in \(Q\).  This
linear term makes the unperturbed state stationary at the imposed
current, but it does not change the quadratic coefficients.  Therefore
the quadratic form can be obtained by expanding the uniform part of
Eq.~\eqref{eq:free_energy} to second order and then dropping the
first-order terms.  These first-order terms vanish because the
unperturbed state satisfies the uniform Usadel equation, the gap
equation, and the fixed-current stationarity condition.

\begin{eqnarray}
\delta^2 {\mathcal G}
&=& \eta^2\ln\frac{T}{\Tc}
 +2\pi\tau\sum_n
 \biggl[
 \frac{\eta^2}{\Omega_n}
 -2\eta c_n \,\alpha_n \nonumber \\
&& + \{ Q^2(c_n^2-s_n^2)+\Omega_n c_n+\Delta s_n\} \alpha_n^2
 \nonumber\\
&&+4Q s_n c_n\, V\alpha_n
 + s_n^2 V^2
 \biggr],
 \label{eq:uniform_quadratic}
\end{eqnarray}
where $c_n :=\cos\theta_n$ and $s_n :=\sin\theta_n$.
For a given \(V\), the lowest-energy perturbation is obtained by minimizing the quadratic energy with respect to the remaining variables, \(\partial_{\alpha_n}\delta^2\mathcal G=0\) and \(\partial_{\eta}\delta^2\mathcal G=0\). The first condition is the linearized Usadel equation, and the second condition is the linearized gap equation. They give 
\begin{eqnarray} 
&&\alpha_n = -\left( \frac{M c_n}{K d_n} +\frac{2Q s_n c_n}{d_n} \right)V, \label{eq:alpha_solution} \\ 
&&\eta = -\frac{M}{K}V , \label{eq:eta_solution} 
\end{eqnarray} 
and, after \(\eta\) and \(\alpha_n\) are eliminated, the remaining quadratic cost in the \(V\) direction is
\begin{equation} 
\delta^2\mathcal G_{\rm min} = \frac{1}{K} \bigl( KL -M^2  \bigr) V^2,
\label{eq:relaxed_quadratic_cost} 
\end{equation} 
Here, 
\begin{eqnarray}
&&d_n:=Q^2(c_n^2-s_n^2)+\Omega_n c_n+\Delta s_n, \\
&&K :=
 \ln\frac{T}{\Tc}
 +2\pi\tau\sum_n
 \left(
 \frac{1}{\Omega_n}
 -\frac{c_n^2}{d_n}
 \right),
 \label{eq:K}
 \\
&&L :=
 2\pi\tau\sum_n
 \left(
 s_n^2
 -\frac{4Q^2s_n^2c_n^2}{d_n}
 \right), \label{eq:L}\\
&&M :=
 4\pi\tau Q\sum_n
 \frac{s_nc_n^2}{d_n},
 \label{eq:M}
\end{eqnarray}
The uniform Usadel equation gives \(d_n=\Omega_n/c_n+Q^2c_n^2>0\), and therefore \(M\ge0\) (\(M>0\) for \(Q>0\)). Moreover, \(K=(\partial h/\partial\Delta)_Q\), where \(h\) is the left-hand side of the gap equation with \(\theta_n\) eliminated by the Usadel equation; stability of the superconducting solution requires this slope to be positive, so \(K>0\).
The vortex-free state is stable while this minimized curvature is positive. It loses local stability when the curvature vanishes, \(KL-M^2=0\). 
This condition defines the stability limit in the spatially uniform sector. We denote the current satisfying this condition by \(J_{\rm v}^{(0)}\). 
At this stage, \(J_{\rm v}^{(0)}\) is only a candidate for the microscopic vortex-entry current: it is the actual \(J_{\rm v}\) only if no spatially nonuniform perturbation becomes unstable at a lower current. 
This point is examined next.

\subsection{Nonuniform perturbations}

Spatially nonuniform perturbations, \(\delta{\bf Q}(x,y)\), \(\delta\Delta(x,y)\), and \(\delta\theta_n(x,y)\), can be treated explicitly by expanding them in the normal modes of the strip. The strip occupies \(-w/2<x<w/2\) and is translationally invariant along \(y\). Here, \(x\) and \(y\) are dimensionless coordinates, and \(w\) is the dimensionless width, all normalized by \(\xi\).
In the self-field-free problem, the vector potential is fixed, so a momentum perturbation comes only from the phase, \(\delta{\bf Q}=\nabla\delta\chi\). Therefore the allowed momentum perturbations are curl-free, \(\nabla\times\delta{\bf Q}=0\). The insulating edges require no normal current through the boundaries; to linear order around the uniform state this gives \(\delta Q_x=0\) at \(x=\pm w/2\). The spectral-angle perturbations satisfy the linearized Neumann condition, \(\partial_x\delta\theta_n=0\), at the edges. The gap perturbation \(\delta\Delta\) has no independent boundary condition in this local Usadel functional, but it can be expanded in the same complete cosine basis. Thus the cosine modes across the strip and Fourier modes along the strip provide a complete mode expansion for all allowed small perturbations, except for the exactly uniform mode already treated above.
We write
\begin{eqnarray} 
{\bf Q} &\to& {\bf Q} + W_{mk} \sqrt{ \frac{2}{\pi w}} \sin\frac{m\pi(x+w/2)}{w}\sin ky \,\hat{\bf x} \nonumber \\
&& + V_{mk} \sqrt{\frac{2-\delta_{m0}}{\pi w}} \cos\frac{m\pi(x+w/2)}{w} \cos ky\,\hat{\bf y}, \\ 
\Delta &\to& \Delta+\eta_{mk} \sqrt{\frac{2-\delta_{m0}}{\pi w}} \cos\frac{m\pi(x+w/2)}{w} \cos ky, \\ 
\theta_n &\to& \theta_n+\alpha_{nmk} \sqrt{\frac{2-\delta_{m0}}{\pi w}} \cos\frac{m\pi(x+w/2)}{w} \cos ky, 
\end{eqnarray} 
Here \(k>0\), although it may be arbitrarily small. 
The orthogonal sine sector along \(y\) gives the same quadratic form and is omitted only to avoid duplication.
The exactly uniform mode, \(m=0\) and \(k=0\), was treated above. 
The \(W_{mk}\) term is absent for \(m=0\). 
For \(m\ge1\), the self-field-free condition \(\nabla\times\delta{\bf Q}=0\) gives $kW_{mk}+(m\pi/w) V_{mk}=0$. 
For the special case \(m=0\), the transverse component is absent, \(W_{0k}=0\), while the longitudinal component \(V_{0k}\) is allowed.

For these nonuniform modes, the fixed-current Legendre term does not affect the quadratic form because it is linear in \(Q_y\) and therefore has no second variation. 
Moreover, the modes with \(k>0\) have zero spatial average of \(\delta Q_y\), while the purely transverse \(k=0\), \(m\ge1\) modes have \(V_{m0}=0\). Thus the Legendre term gives no first-order contribution for these modes. Only the exactly uniform mode changes the spatially averaged \(Q_y\), and that mode was treated above.

With this normalization, the quadratic variation for each mode takes the
same form as in the uniform calculation:
\begin{eqnarray} 
\delta^2\G_{mk} 
&=& \eta_{mk}^2\ln\frac{T}{T_c} +2\pi\tau\sum_n \biggl[ \frac{\eta_{mk}^2}{\Omega_n} 
-2\eta_{mk}c_n\alpha_{nmk} \nonumber \\
&& +(d_n+p_{mk}^2)\alpha_{nmk}^2 
 +4Q s_n c_n V_{mk} \alpha_{nmk} \nonumber \\
&& +s_n^2 (W_{mk}^2+V_{mk}^2 ) \biggr], 
\end{eqnarray} 
where \(p_{mk}^2:=(m\pi/w)^2+k^2\).  For \(m=0\), the transverse
momentum mode is absent; in the formula above this is understood as
\(W_{0k}=0\).

For a given \(V_{mk}\), the lowest-energy values of \(\alpha_{nmk}\) and \(\eta_{mk}\) are determined by the linearized Usadel equation and the linearized gap equation. After these variables are eliminated, the minimized second variation is
\begin{eqnarray} 
\delta^2\G_{mk}^{\rm min} = \left( {\mathcal L}_{mk} -\frac{{\mathcal M}_{mk}^2}{{\mathcal K}_{mk}} \right)V_{mk}^2 +S W_{mk}^2 . 
\end{eqnarray} 
Here, the eliminated fields and coefficients are given by
\begin{eqnarray} 
&&\alpha_{nmk} = \frac{ c_n\eta_{mk}-2Q V_{mk}s_nc_n }{ d_n+p_{mk}^2 } , \\
&&\eta_{mk} = -\frac{{\mathcal M}_{mk}}{{\mathcal K}_{mk}}V_{mk}, 
\end{eqnarray} 
and
\begin{eqnarray} 
{\mathcal K}_{mk} &:=& \ln\frac{T}{T_c} +2\pi\tau\sum_n \left( \frac{1}{\Omega_n} -\frac{c_n^2}{d_n+p_{mk}^2} \right), \\ 
{\mathcal L}_{mk} &:=& 2\pi\tau\sum_n \left( s_n^2 -\frac{4Q^2s_n^2c_n^2}{d_n+p_{mk}^2} \right), \\
{\mathcal M}_{mk} &:=& 4\pi\tau Q\sum_n \frac{s_nc_n^2}{d_n+p_{mk}^2}. 
\end{eqnarray} 
For any nonuniform mode with \(p_{mk}^2>0\), the denominator \(d_n+p_{mk}^2\) is larger than \(d_n\). Therefore \({\mathcal K}_{mk}>K\), \({\mathcal L}_{mk}\ge L\), and \(0\le{\mathcal M}_{mk}\le M\) on the stable uniform superconducting solution. 
For \(Q>0\), the last two inequalities are strict, and the minimized curvature
\({\mathcal L}_{mk}-{\mathcal M}_{mk}^2/{\mathcal K}_{mk}\)
is strictly larger than the uniform curvature \(L-M^2/K\).
At \(Q=0\), the two longitudinal curvatures are equal but positive.
In addition, modes with \(m\ge1\) have the positive
transverse-momentum contribution \(S W_{mk}^2\). 
Therefore no spatially nonuniform mode can become unstable before the uniform mode in the ideal zero-field, self-field-free strip.

The purely transverse modes \((m,k)=(m,0)\) with \(m\ge1\) are not included in the \(k>0\) continuous Fourier normalization above. For these modes, the curl-free condition gives \(V_{m0}=0\). After the spectral-angle perturbations are eliminated, the amplitude sector has the positive coefficient \({\mathcal K}_{m0}>K>0\), while the phase sector contains the positive transverse-momentum cost \(S W_{m0}^2\). Therefore these modes also remain stable up to the uniform instability.

\subsection{Vortex-entry current}

The analysis above shows that, among the allowed perturbations of the ideal zero-field, self-field-free strip, the uniform mode sets the stability limit and therefore determines the vortex-entry current: \(J_{\rm v}= J_{\rm v}^{(0)} \). 
We now return to this uniform mode. As shown above, the minimized quadratic energy cost \(\delta^2\G_{\min}\) vanishes when \(KL-M^2=0\), where \(K\), \(L\), and \(M\) are defined in Eqs.~(\ref{eq:K})--(\ref{eq:M}). This curvature is directly related to the slope of the self-consistent current--momentum curve \(J(Q)\). 
Along the self-consistent uniform solution at nearby \(Q\), which is the bottom of the energy valley obtained after \(\Delta\) and \(\theta_n\) are relaxed, 
a change \(Q\to Q+V\) is accompanied by \(\Delta\to\Delta+(d\Delta/dQ)V\) and \(\theta_n\to\theta_n+(d\theta_n/dQ)V\). 
Thus \(d\Delta/dQ=\eta/V\) and \(d\theta_n/dQ=\alpha_n/V\), and these ratios are precisely the responses obtained in Eqs.~(\ref{eq:alpha_solution}) and (\ref{eq:eta_solution}). Differentiating \(J(Q)\) then gives
\begin{eqnarray} 
\frac{dJ}{dQ} 
&=& 4\sqrt{\pi}\,\tau \Js \sum_n \left[ s_n^2 +2Qs_nc_n\frac{d\theta_n}{dQ} \right]\nonumber \\
&=& \frac{2\Js}{\sqrt{\pi} K} (KL-M^2) 
\label{eq:dJdQ_1} 
\end{eqnarray} 
Equation~\eqref{eq:dJdQ_1} is the central identity of this work. It shows that the barrier-disappearance condition \(KL-M^2=0\) and the depairing condition \(dJ/dQ=0\) are not two different conditions that happen to be satisfied at the same current. They are algebraically equivalent conditions for the same loss of stability of the vortex-free state. Therefore, for an ideal homogeneous narrow thin-film strip at any temperature \(0<T<\Tc\),
\begin{eqnarray}
J_{\rm v}(T)=J_{\rm dp}(T).
\end{eqnarray}
The temperature dependence of \(J_{\rm dp}\) has been extensively studied (see, e.g., Refs.~\cite{Kubo_2020,Kubo_2022, Kubo_2025} and references therein).

\section{Conclusion}

We have identified the microscopic counterpart of the Pearl--London barrier-disappearance current. The result is stronger than a numerical equality between two independently defined currents. The microscopic condition for disappearance of the vortex-entry barrier and the depairing condition are identical: both identify the spinodal at which the vortex-free current-carrying state loses local stability. Thus, for an ideal homogeneous strip at zero applied field, with self-field effects neglected, $J_{\rm v}(T)=J_{\rm dp}(T)$ over the full temperature range $0<T<T_c$.

Nonideal systems may involve additional stability mechanisms that lie outside the equivalence established here~\cite{Buzdin,Yu,Vodolazov_2003,Kubo_2015,Lechner,Kerman,Hortensius,Clem_Berggren,Clem_Mawatari,Baghdadi,Jonsson,Bartolf,Yamashita,Bulaevskii,Jahani,Kubo_2023}. Extending the present microscopic framework to such systems remains a subject for future work.

\begin{acknowledgments}
This work was supported by JSPS KAKENHI under Grant Nos. JP25K01610, JP25K23386, JP26K03209, and JP26K00665. 
The idea for this work emerged during my three-year paternity leave from 2021 to 2024. I am deeply grateful to everyone who supported me during that period, which was made possible by the Act on Childcare Leave of Japan~\cite{ikuji}.
\end{acknowledgments}

\end{document}